\documentclass[fleqn,twoside]{article}
\usepackage{amsmath,espcrc2,graphics}

\usepackage{graphicx}

\def\beq{\begin{equation}}
\def\eeq{\end{equation}}
\def\beqn{\begin{eqnarray}}
\def\eeqn{\end{eqnarray}}
\def\nonu{\nonumber}
\def\MS{\overline{\rm MS}}

\title{Jet and Di-Jet production in Photon-Photon collisions}

\author{L. Bertora\address[DiFI]{Dipartimento di Fisica, Universit\'a di Genova and \\
Istituto Nazionale di Fisica Nucleare, Sezione di Genova,\\
        Via Dodecaneso 33, 16146 Genova, Italy}%
        \thanks{bertora@fisica.unige.it}
       }
       
\begin{document}
\begin{abstract}
We present next-to-leading order QCD predictions for single-inclusive and
di-jet observables, relevant to the collisions of two on-shell photons,
 obtained with a recently completed computation based
upon the subtraction method. We compare these predictions with available
theoretical results, with LEP data, and we discuss the various uncertainties
which affect them.
\vspace{1pc}
\end{abstract}

\maketitle

\section{Introduction}
\setcounter{footnote}{0}

During the past two years different analyses \cite{wengler,L3pro,pablo,delphi} of jet
and di-jet production from real photons scattering at LEP have been presented. 
A unique computation it is available in the literature \cite{kkk,kk}
 at NLO for this process.
This computation is based on the slicing method to cancel
 analytically virtual and infrared\footnote{We use the word ``infrared''
 both for the properly soft effects and for the collinear ones.} 
divergences. 
This technique relies on an approximate treatment of the phase space
and of the matrix elements, which induces cancellations between large numbers.
 Hence in particular regions of  the phase space it is possible that these 
cancellations are difficult to achieve numerically.
To avoid this problem (it could be substantial for specific
exclusive observables which have been measured \cite{wengler}), it seems appropriate
 to use an exact technique to handle the infrared divergences. 
We use the subtraction method in the formulation of ref. \cite{stefanoks,steparton}.
This method treats exactly both the phase space and the matrix elements, thus
ensuring a better behaved numerical result. 
We present a complete computation of jet and di-jet cross-sections 
and we compare our theoretical predictions with previous results
and with experimental data. 
 
\section{Some remarks about subtraction method}

To compute a generic  physical observable in hadronic physics it is necessary
 to implement the factorization theorem
in a computer code. Thus we have to deal with unphysical large oscillations
(due to the cancellation of the infrared divergences following the factorization
prescription) and the computer precision. The solution of this problem it is to
remove divergences analitycally and then to use a numerical method
 (typically a MC algorithm) to integrate the finite reminders. 
In this section we briefly describe a simplified 1-D model \cite{stefanoye} to
outline the differences between the subtraction and the slicing method.

\subsection{Application of the slicing and the subtraction method to a simple case}

Let us consider the integral,
\beqn
\langle F \rangle=\frac{1}{2\epsilon} \int_0^1{dx \delta(1-x)F(x)}+ \nonumber\\ 
 \int_0^1{dx (1-x)^{-1-2\epsilon}F(x)},
\label{1d}
\eeqn
in the limit of $\epsilon\rightarrow 0$, where $F$ is any regular function.
 This is a simplified situation
of what happens to cancel out the soft emission (the second term in (\ref{1d}))
 and the virtual correction (the first term) divergent terms when we compute a generic infrared safe observable $F$;
$\epsilon$ represents the (dimensional) infrared regulator.
The first integral is trivial, the second one is not; in general,
 we can not compute it analytically.
To remove the $\epsilon$ pole we need, at least, to express explicitly the divergent part
of the second integral.
If we use the slicing strategy  we rewrite the second term of (\ref{1d}) as follows:
\beqn
\int_0^1{dx \frac{F(x)}{(1-x)^{1+2\epsilon}}}= 
\int_0^{1-\delta}{dx \frac{F(x)}{(1-x)^{1+2\epsilon}}}+ \nonumber\\
\int_{1-\delta}^1{dx \frac{F(x)}{(1-x)^{1+2\epsilon}}},\text{  } 0<\delta\leq 1.
\label{sli}
\eeqn
It is clear that the only divergent term is in the second piece on the RHS,
and if we perform a Taylor expansion around $\delta=0$, we obtain the $\epsilon$ pole we need.
Thus singular terms in $\epsilon$ in Eq. (\ref{1d}) cancel out, and we obtain
\beqn
\langle F \rangle= 
\int_0^{1-\delta}{dx \frac{F(x)}{1-x}}+F(1)\log\delta +O(\delta,\epsilon).
\label{sliresult}
\eeqn
This formula is obviously finite if we set $\epsilon=0$ whatever $\delta$ we choose, thus 
we can perform  the numerical integration. Formally the $\delta$ dependence of 
(\ref{sliresult}) is due only to the Taylor approximation we have done and it is of first order
 because the log$\delta$ term is canceled by an analogous factor coming from the upper integration 
 limit in the (\ref{sliresult}). Then it is clear that the smaller $\delta$, the less accurate 
the numerical cancellations.\\
On the other hand if we use the subtraction technique we can rewrite (\ref{sli}) as
\beqn
\int_0^1{dx \frac{F(x)-F(1)\theta(x-1+x_c)}{(1-x)^{(1+2\epsilon)}}}+ \nonu \\
F(1)\int_0^1{dx \frac{\theta(x-1+x_c)}{(1-x)^{(1+2\epsilon)}}}, \text{  }
0<x_c\leq 1. 
\label{sub}
\eeqn
This time we have added and removed the same term, essentially the singular behavior of 
the integrand, in such a way that the divergent piece is isolated in the second factor as before.
But this time we can perform exactly the integration of the divergent part. After removing
the $\epsilon$ pole and we obtain
\beqn
\langle F \rangle= \int_0^{1}{dx \frac{F(x)-F(1)\theta(x-1+x_c)}{1-x}}+ \nonu \\
 F(1)\log x_c.
\label{subresult}
\eeqn
The logarithm term the equation (\ref{subresult}) is formally identical to the 
one in (\ref{sliresult}). However, here no large number cancellations
are present because we are not forced to set $x_c \approx 1$. Thus the results
 from (\ref{subresult}) are exact and free of large fluctuations, 
but a good numerical integration is still necessary to obtain accurate results.

\section{Structure of the computation}

As is well known, in real photon-photon scattering photons can interact
 directly or hadronically. Thus the interaction is conventionally divided in three classes of Feynman diagrams:
double resolved class, where the two photons interact by their hadronic components,
the single resolved class, where only one photon interact by its hadronic component,
and the direct class, where the photons themselves are the initial states
 of the hard process. As we will explicitly show each contribution has no
physical meaning but only the sum correspond to an observable.
For the first and second class we   
have modified the code \cite{gioste} used to compute photon-hadron jet observables.  
As far as direct contribution is concerned we have recalculated and checked with \cite{kkk,Aurenche} all the relevant matrix elements,
which appear in the computation at NLO ($\alpha^2\alpha_S$ order).
Finally we have implemented the third class of diagrams in a computer code, analogue to the previous two codes.
By summing the results from the three codes, we can obtain, using the subtraction method,
any infrared finite jet and di-jet observable at NLO in real photon collisions.
The details of the computation can be found in \cite{us}. 

\subsection{Consistency checks}

\begin{figure}[htb]
\includegraphics*[scale=0.5]{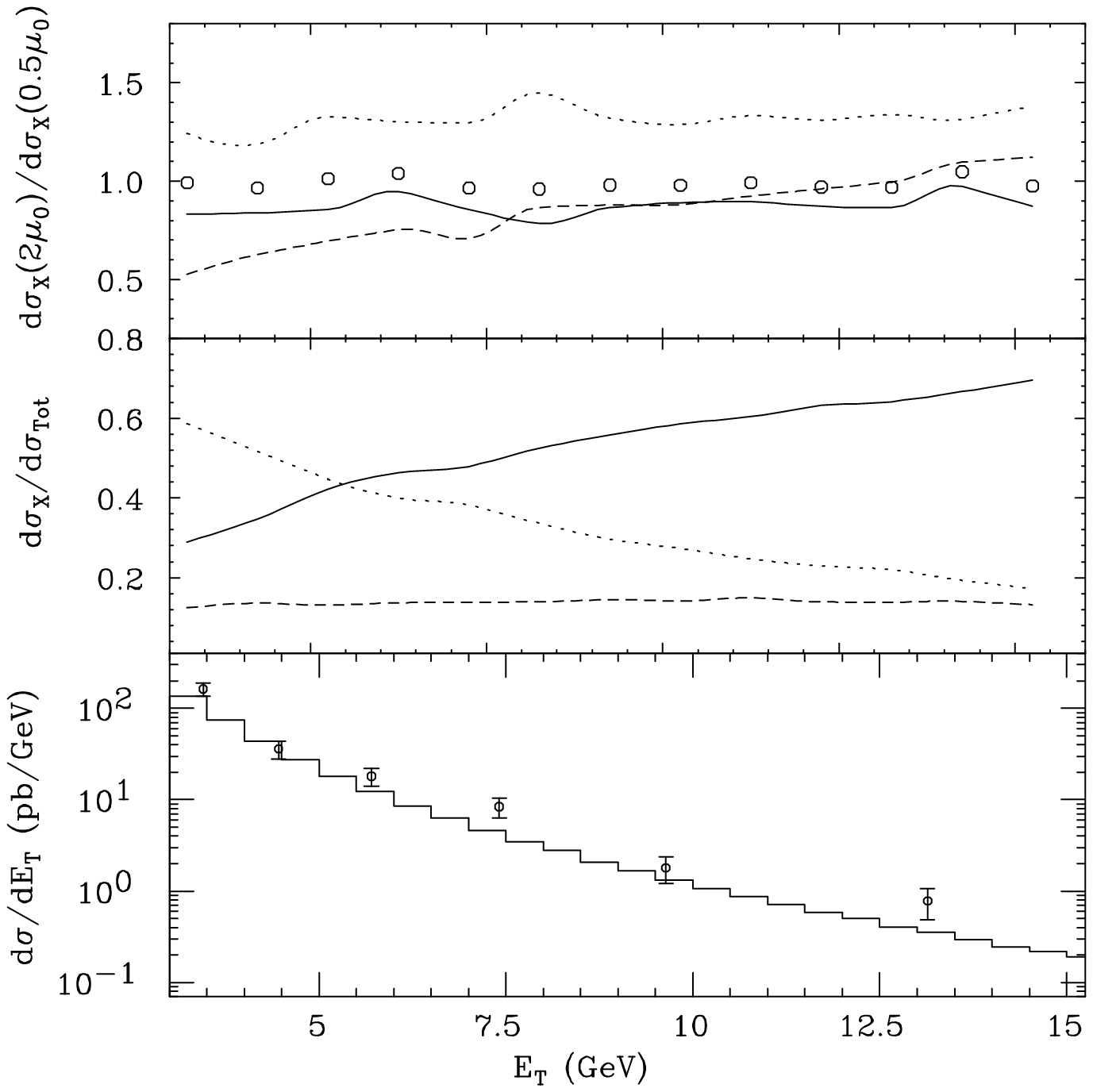}
\label{O130fig}
\caption{}
\end{figure}

\begin{figure}[htb]
\includegraphics*[scale=0.5]{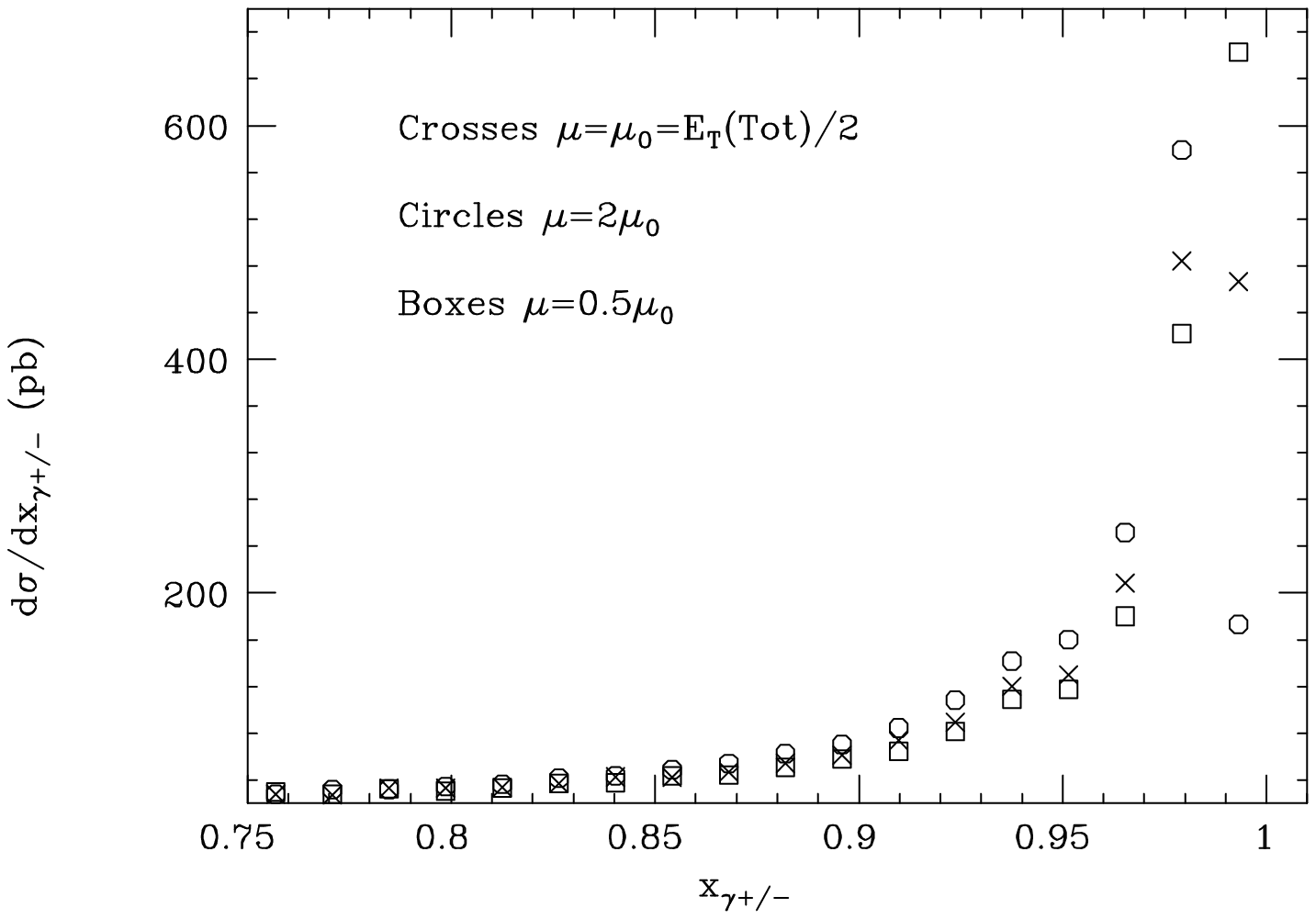}
\label{xscalefig}
\caption{}
\end{figure}

We start by presenting the comparison of our theoretical predictions
 with the LEP OPAL data \cite{OPAL1997}.
 In the lower panel of Figure~1 the
differential inclusive one jet cross-section is shown. 
As one can see our computation (we use in Figure~1 $\MS$ scheme,
 $\mu_0=\mu_{R}=\mu_{F}=E_T^{Tot}/2$,\footnote{Note that the half of total transverse energy 
of the event ($E_T^{Tot}$) is equal to $E_T$-jet at LO.} GRV-HO 
photon density functions, $\Lambda_{QCD}=130$ MeV; see \cite{OPAL1997}
 for the jet definition, for the cuts and for
the  kinematic) is in decent agreement with experimental data.
In the central panel the ratios of the double resolved (dots), single resolved (dashed),
direct (solid) contributions and of the sum are plotted.
Our computation agrees  with the previous one \cite{kkk}.
Let us study the consistency of our computation, for example by changing the 
factorization scale ($\mu_{F}=2\mu_0,0.5\mu_0$).
In fact if we change the factorization scale,
 upper panel of Figure~1,
 we obtain the same theoretical prediction (circles) but the double,
 single resolved and direct contributions change.
Because 
each class of Feynman diagrams is separately divergent while the sum is finite,
in order to 
perform the computation we had subtracted divergences analytically from each class,
adding (or removing) an infinite term. Such a procedure depends on the regularization
we use, i.e. on the factorization scale and scheme.		
Hence changing regularization the partial results change, the total does not\footnote{Up to uncontrolled higher-order terms, which appear to be negligible in this case.}.
In conclusion Figure~1 represents a good consistency check of our computation and shows that the separation of Feynman graphs at NLO is physically meaningless. \\
To distinguish between the two types of interactions of the photons,
it is customary to introduce the couple of variables
\beq
x^{\pm}_{\gamma}=\frac{\sum_{Jet=1,2}(E^{J}\pm p^{J}_z)}
               {\sum_{Hadrons}(E^{H}\pm p^{H}_z)},
\label{x+-}
\eeq
where the sum in the numerator runs over the two hardest jets of the event.
The meaning of the definition (\ref{x+-}) it is clear: in fact if we
consider the Born process in the direct case, we have only two
back-to-back jets  and no hadronic remnants, which implies  $x^{\pm}_{\gamma}=1$.
Thus generally the closer $x^{\pm}_{\gamma}$ to one, the more the cross section is 
dominated by the direct contribution. Elsewhere the process is dominated by
 resolved interactions. We are able to distinguish the behavior of
each photon (coming from the electron or the positron), and then we can 
study the single resolved case too.
It is clear that we need a $x_{\gamma}^{sep}$ value to separate (in an arbitrary manner)
 between the hadronic and the direct contribution: we will use  $x_{\gamma}^{sep}=0.75$, 
the same value used by OPAL,
suggested by MonteCarlo simulations \cite{wengler}.
Therefore we consider three regions on the $x^{\pm}_{\gamma}$-plane, corresponding to 
the three types of interactions mentioned before. We expect that the
 cross-section in  the ``hadronic'' region to be the most sensitive to
the PDF choice. Indeed this is the case.
Moreover we can predict the differential 
cross-sections with respect to $x^{\pm}_{\gamma}$. To this aim it is interesting
to check what happens if we change again the factorization scale. 
In Figure~2 we plot $d\sigma/dx^{\pm}_{\gamma}$ for three different scales in the region 
$x^{\pm}_{\gamma}>0.75$. For $x^{\pm}_{\gamma}<0.80$ the curves are indistinguishable,
 but it is evident that close to one we obtain unstable results,
 strongly depending on the factorization scale choice.

\section{IR-Sensitive regions} 

We can deduce physical observations from these consistency tests.
In fact if we look at the hadronization corrections\footnote{Here, the hadronization correction is
 defined as the ratio between the MC cross-sections at the parton and hadron levels.}
 taken from \cite{wengler,hepdata}
 to $d\sigma/dx^{\pm}_{\gamma}$ in the same region 
we observe that they are between 40\% and 260\%.
Hence the conclusion is that the perturbative computation at NLO fails in this region.
To understand this fact let us consider the relationship between $x^{\pm}_{\gamma}$ and the 
fraction of energy $(1-x_\gamma^{1,2})$ of the emitter parton carried out by a soft gluon emitted
in the final state of the direct process.
It is simple to show that in a three-jet configuration 
\beq
x_\gamma^{1,2}\rightarrow 1 \Rightarrow x^{\pm}_{\gamma} \rightarrow 1.
\eeq
Thus the ``dangerous region'' showed in Figure~2 corresponds to a soft gluon emission kinematic,
where the exponentiation and the resummation to all order of the large logarithms are necessary to 
obtain sensible predictions.
In particular the dominant term of the perturbative differential
 cross-section in the limit $x_\gamma^{\pm}\rightarrow 1$ is roughly 
\beq
\frac{d\sigma}{dx_{\gamma}^{\pm}}\approx \frac{1}{1-x_{\gamma}^{\pm}}-
\frac{\theta(x_{\gamma}^{\pm}-1)}{1-x_{\gamma}^{\pm}}.
\label{x12div}
\eeq
The expression (\ref{x12div}) explicitly indicates the finiteness 
(i.e. the infrared safeness) of the cross-section we expect from
jet definitions, but because infrared divergence is canceled badly
(exactly for $x_\gamma^\pm=1$) large logarithms appear. 
Thus the comparison of the data in this region with NLO predictions cannot be satisfactory.
In other words one has to renounce to compare the exclusive observable $d\sigma/dx^{\pm}_{\gamma}$ in this region with fixed order computations. 
To avoid this problem one can use more inclusive quantities \cite{wengler,busseypro1}, obtained
by integrating  $d\sigma/dx^{\pm}_{\gamma}$ over $x_\gamma^{\pm}\ge 0.8$. This improves the agreement between data and 
predictions.\\  
We also point out that there is another region in the phase space which can introduce large 
logarithmic corrections to NLO computations.
It is known that di-jet observables are sensible to the choice of
the energy cuts on the phase space, in particular it is shown \cite{gioste}
that symmetric cuts over the two jets transverse energy imply large logarithmic 
corrections to certain observables.
Thus to avoid this effect OPAL collaboration has chosen \cite{wengler} 
asymmetric cuts on the $E_T^{Jet}$-plane defined as
\beqn
\nonumber
\overline{E}_T=(E^1_T +E^2_T)/2 \ge\text{ 5 GeV,}\\
0\leq \phi=\frac{\left|E^1_T -E^2_T\right|}{E^1_T +E^2_T} <
 \frac{1}{4}. 
\eeqn
Even if these cuts are free from the troubles we cited before, we point out that other 
problems could arise for NLO predictions. Indeed,
NLO computation imposes $\phi \leq 1/3$, and setting the $\phi$-limit near this
value could be problematic. Furthermore if the $\phi$-limit goes to zero 
infrared effects arise. In fact one should expect that the cross-section goes to zero if
the phase space becomes small, but as Figure~3 shows this is not the case at NLO.

\begin{figure}[htb]
\includegraphics*[scale=0.5]{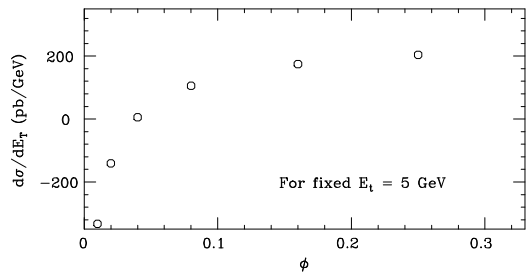}
\label{philimitfig}
\caption{Variation of the cross-section with the $\phi$-cut at fixed $\overline{E}_T=5$ GeV}
\end{figure}

The explanation of this is simple: if $\phi\rightarrow 0$ then there is not enough phase
space to emit a soft gluon to cancel out properly virtual divergences.
Therefore large logarithms of $\phi$ \cite{us}
 become dominant when $\phi\rightarrow 0$.
Thus even if in principle OPAL choice \cite{wengler} could be dangerous in practice
it is far from the ``dangerous regions''.
            
\section{Phenomenological results}

\begin{figure}[htb]
\includegraphics*[scale=0.48]{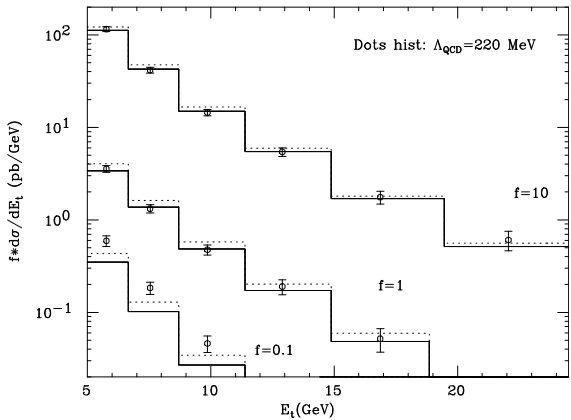}
\label{weninclusive}
\caption{Differential di-jet cross-section measured by OPAL compared with our prediction}
\end{figure}
In this section we compare our results to recent data.
In Figure~4 our theoretical predictions are compared to OPAL measurements
\cite{wengler} (see that paper for the kinematics and the cuts) 
for di-jet cross-sections are plotted . The three series of curves 
correspond respectively to the whole $x^{\pm}_{\gamma}$ region, either 
$x^{+}_{\gamma}> 0.75$
or $x^{-}_{\gamma}>0.75$, and $x^{\pm}_{\gamma}<0.75$. We use for the central 
theoretical prediction GRV-HO in DIS scheme with  $\mu_0=\mu_{R}=\mu_{F}=E_T^{Tot}/2$
and $\Lambda_{QCD}=130$ MeV. The NLO predictions, for all the observables we present,
 are divided by the factor $(1+\delta_H)$,
the hadronization corrections taken from \cite{wengler,hepdata}. These terms are less than 10\% for
$\overline{E}_T$ grater than 10 GeV, but rise to about 20\% towards small $\overline{E}_T$.
Plots show a very good agreement between QCD predictions and data for the most 
inclusive observable and in the single resolved region, whereas the agreement does not
 seem satisfactory in the double resolved region. Even if we consider the theoretical
errors due to the scale and the PDF choice or $\Lambda_{QCD}$, as shown in Figure~4, we are not able to adsorb
discrepancy in the theoretical uncertainties. Furthermore the situation is worse if 
we consider  $d\sigma/dx^{\pm}_{\gamma}$ in this region, NLO prediction cannot satisfactory describe the experimental data.
The same discrepancies affect the comparison between data and the previous theoretical computation \cite{wengler}. 
Thence we conclude that more experimental and theoretical studies are needed.

\begin{figure}[htb]
\includegraphics*[scale=0.5]{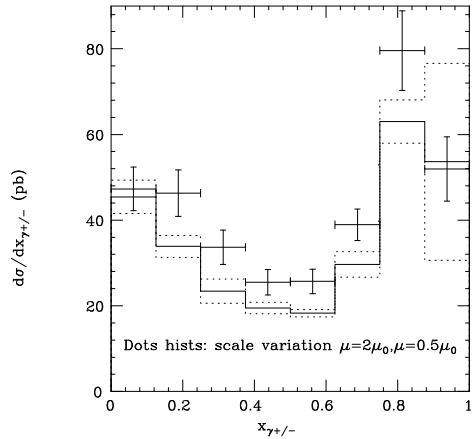}
\label{wenexclusive}
\caption{Differential di-jet cross-section with respect to $x^\pm_\gamma$, $5<\overline{E}_T<7$ GeV.}
\end{figure}

If we compare (Figure~5) $d\sigma/dx^{\pm}_{\gamma}$ with the data for  $0\le x^{\pm}_{\gamma}\le 1$ and $5<\overline{E}_T<7$ GeV (where the ``resolved'' component is dominant) we have only a partial agreement. We note that the correspondence we find
for  $x^{\pm}_{\gamma}> 0.8$ is not particularly meaningful. In fact, 
as we have discussed in Section~3 the result is not stable under variation of the factorization scale.
On the other hand in the energy range $7<\overline{E}_T<11$ GeV data agree well with the theoretical curves outside
the ``dangerous region'', i.e. for $x_{\gamma}^{\pm}<0.8$.   
Other quantities measured by OPAL have been computed and good agreement has been found
 with the theoretical predictions.

\begin{figure}[htb]
\includegraphics*[scale=0.48]{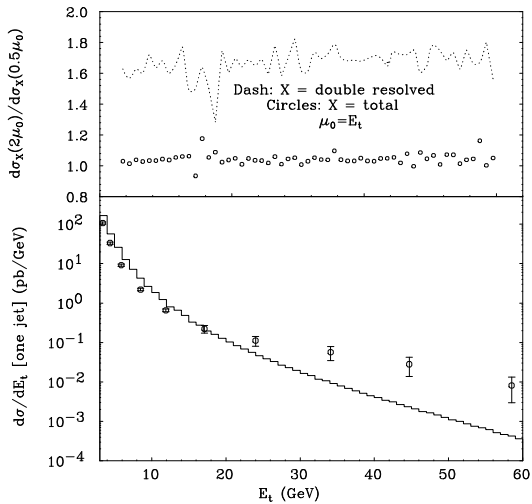}
\label{l3fig}
\caption{Differential single jet cross-section measured by L3: Preliminary Data}
\end{figure}

Recently, an L3 analysis has been presented \cite{L3pro}, for the inclusive  single jet
$d\sigma/dE_T$ on a very extended $E_T$ range ($E_T^{max}\approx$ 60 GeV).
In Figure~6 we present
the comparison between our prediction and L3 preliminary data. As one can see
 it is possible to divide the momentum range in two regions.
For momenta less than 15 GeV the theoretical curve has the right shape to fit
the data but it has the wrong normalization (approximately it is a factor two higher).
For large momenta data and theoretical prediction are clearly in disagreement.
The discrepancy is at places of one order of magnitude, which cannot be accounted for by the theoretical uncertainties.
In the upper panel of Figure~6 the scale dependence of the theoretical prediction (circles) is shown, and similarly to the case of
Figure~1 the variation is very small. Dotted curve re-shows the scale dependence of the ``hadronic'' contribution. 
\section{Conclusion}

We have implemented a computer code which allows to compute jet and di-jet observables
at NLO using the subtraction method. Our results are in agreement with the 
previous theoretical predictions obtained with the slicing method \cite{kkk}.
The comparison of our results with OPAL data \cite{wengler} is satisfactory
for inclusive observables, whereas the comparison with L3 data \cite{L3pro} display sizable disagreement in the large $E_T$ region.\\
Finally, we have analyzed some theoretical difficulties to describe
particular exclusive observables measured in \cite{wengler}.
The theoretical uncertainties for inclusive quantity are under control.

\end{document}